\documentclass[twocolumn]{aastex6}
\usepackage{amsmath}

\newcommand{\Eqn}[1]{Equation~\ref{#1}}
\newcommand{\Fig}[1]{Figure~\ref{#1}}
\newcommand{\Tab}[1]{Table~\ref{#1}}
\newcommand{\Sect}[1]{Section~\ref{#1}}
\newcommand{\App}[1]{Appendix~\ref{#1}}

\DeclareMathOperator{\sech}{sech}
\newcommand{\spt}{\ensuremath{{\cal S}}}

\newcommand{\dd}{\mathrm{d}}

\shorttitle{Brown Dwarfs in the Galactic Disk}
\shortauthors{Ryan et al.}
\begin{document}

\title{The Effect of Atmospheric Cooling on the Vertical Velocity Dispersion and Density Distribution of Brown Dwarfs}\footnote{Support for program \#13266 was provided by NASA through a grant from the Space Telescope Science Institute, which is operated by the Associations of Universities for Research in Astronomy, Incorporated, under NASA contract NAS 5-26555.}

%\correspondingauthor{Russell Ryan}
\email{rryan@stsci.edu}

\author{Russell E.~Ryan Jr.\altaffilmark{1},
  Paul A.~Thorman\altaffilmark{2},
  Sarah J.~Schmidt\altaffilmark{3},
  Seth H.~Cohen\altaffilmark{4},
  Nimish P.~Hathi\altaffilmark{1,5},
  Benne W.~Holwerda\altaffilmark{6,7},
  Jonathan I.~Lunine\altaffilmark{8},
  Nor Pirzkal\altaffilmark{1},
  Rogier A.~Windhorst\altaffilmark{4}, \and 
  Erick Young\altaffilmark{9}}
\altaffiltext{1}{Space Telescope Science Institute, 3700 San Martin Dr., Baltimore, MD 21218}
\altaffiltext{2}{Haverford College, Departments of Physics and Astronomy, 370 Lancaster Ave., Haverford, PA 19041}
\altaffiltext{3}{Leibniz-Institut f\"{u}r Astrophysik --- Potsdam (AIP), An der Sternwarte 16, 14482 Potsdam, Germany}
\altaffiltext{4}{Arizona State University, School of Earth and Space Exploration, P.O. Box 871404, Tempe, AZ 85281-1404}
\altaffiltext{5}{Aix Marseille Universit\', CNRS, LAM (Laboratoire d'Astrophysique de Marseille) UMR 7326, 13388, Marseille, France}
\altaffiltext{6}{Leiden Observatory, Leiden University, Niels Bohrweg 2, NL-2333 CA Leiden, the Netherlands}
\altaffiltext{7}{University of Louisville, Department of Physics and Astronomy, Louisville, KY 40292} 
\altaffiltext{8}{Department of Astronomy and Carl Sagan Institute, Space Sciences Building Cornell University, Ithaca, NY 14853}
\altaffiltext{9}{NASA Ames SOFIA Science Center, N211, Mountain View, CA 94043}

%\author{Russell E. Ryan Jr.}
%\affiliation{Space Telescope Science Institute, 3700 San Martin Dr., Baltimore,% MD 21218}
%
%\author{Paul A. Thorman}
%\affiliation{Haverford College, Departments of Physics and Astronomy, 370 Lanca%ster Ave., Haverford, PA 19041}
%
%\author{Sarah J. Schmidt}
%\affiliation{Leibniz-Institut f\"{u}r Astrophysik --- Potsdam (AIP), An der Ste%rnwarte 16, 14482 Potsdam, Germany}
%
%\author{Seth H. Cohen}
%\affiliation{Arizona State University, School of Earth and Space Exploration, P%.O. Box 871404, Tempe, AZ 85281-1404}
%
%\author{Nimish P.~Hathi}
%\affiliation{Space Telescope Science Institute, 3700 San Martin Dr., Baltimore,% MD 21218}
%\affiliation{Aix Marseille Universit\', CNRS, LAM (Laboratoire d'Astrophysique %de Marseille) UMR 7326, 13388, Marseille, France}%%%
%
%\author{Benne W.~Holwerda}
%\affiliation{Leiden Observatory, Leiden University, Niels Bohrweg 2, NL-2333 CA% Leiden, the Netherlands}
%\affiliation{University of Louisville, Department of Physics and Astronomy, Lou%isville, KY 40292} %
%
%\author{Jonathan I.~Lunine}
%\affiliation{Department of Astronomy and Carl Sagan Institute, Space Sciences Building Cornell University, Ithaca, NY 14853}
%
%\author{Nor Pirzkal}
%\affiliation{Space Telescope Science Institute, 3700 San Martin Dr., Baltimore, MD 21218}
%
%\author{Rogier A.~Windhorst}
%\affiliation{Arizona State University, School of Earth and Space Exploration, P.O. Box 871404, Tempe, AZ 85281-1404}
%
%\author{Erick Young}
%\affiliation{NASA Ames SOFIA Science Center, N211, Mountain View, CA 94043}

\begin{abstract}

We present a Monte Carlo simulation designed to predict the vertical
velocity dispersion of brown dwarfs in the Milky Way.  We show that
since these stars are constantly cooling, the velocity dispersion has
a noticeable trend with spectral type.  With realistic assumptions for
the initial-mass function, star-formation history, and the cooling
models, we show that the velocity dispersion is roughly consistent
with what is observed for M dwarfs, decreases to cooler spectral
types, and increases again for the coolest types in our study
($\sim$T9).  We predict a minimum in the velocity dispersions for L/T
transition objects, however the detailed properties of the minimum
predominately depend on the star-formation history.  Since this trend
is due to brown dwarf cooling, we expect the velocity dispersion as a
function of spectral type should deviate from constancy around the
hydrogen-burning limit.  We convert from velocity dispersion to
vertical scale height using standard disk models, and present similar
trends in disk thickness as a function of spectral type.  We suggest
that future, wide-field photometric and/or spectroscopic missions may
collect sizable samples of distant ($\sim\!1$~kpc) of dwarfs that span
the hydrogen-burning limit.  As such, we speculate that such
observations may provide a unique way of constraining the average
spectral type of hydrogen-burning.

\end{abstract}


\section{Introduction} \label{sec:intro}

Brown dwarfs  are substellar objects  that lack the necessary  mass to
sustain       hydrogen        fusion       in        their       cores
\citep[e.g.][]{kumar62,hayashi}.  Hence, these  objects are constantly
cooling throughout their  lifetimes \citep[e.g.][]{bhl89}, which leads
to  the known  mass-age  degeneracy:  a brown  dwarf  detected with  a
specific temperature could have a range of ages, depending on its mass
\citep[e.g.][]{bc96,burrows97}.

Disk kinematics are powerful probes  of the structure and evolution of
the Milky  Way, but also  provide constraints  on the process  of star
formation.  It  is commonly  accepted that stars  are formed  in giant
molecular clouds that  occupy a thin Galactic disk,  then diffuse over
time   by    successive   interactions   with   the    disk   material
\citep[e.g.][]{spitz51}.   This so-called  {\it disk  heating} can  be
used as a Galactic chronometer,  which makes stellar kinematics a weak
tracer  of  stellar  population  age  \citep[e.g.][]{wiel77}.   Nearby
populations  of early-M  dwarfs are  observed to  have {\it  kinematic
  ages}  of $\sim\!3$~Gyr  \citep[e.g.][]{reid02,rb09}, but  extending
this  chronology   to  the  substellar  population   is  not  trivial,
particularly  if   their  formation  process  differs   from  that  of
main-sequence stars.   \citet{luhman12} review the  proposed formation
scenarios for  brown dwarfs and  the various effects on  their initial
kinematics  and present-day  spatial  distributions.  Early  formation
models speculated  that brown  dwarfs are preferentially  ejected from
their  birthsites  \citep[e.g.][]{rc01},  which should  enhance  their
initial   velocities  and   imply  kinematic   ages  older   than  the
main-sequence population. Most kinematic  and spatial distributions of
cluster brown dwarfs seem to rule out the ejection scenario, and favor
models that have initial  kinematics comparable to main-sequence stars
\citep[e.g.][]{wb03,j06,luhman06,parker11}.    Moreover   disk   brown
dwarfs are  reported with  a range  of kinematic  ages, in  same cases
younger than  M-dwarfs \citep[e.g.][]{zo07, faherty09,sjs10}  or older
\citep[e.g.][]{seif09,blake10,burg15}.  Whether the  lack of empirical
consensus stems from  some observational effect (such  as small sample
size) or points to new  astrophysics (such as formation mechanisms) is
not readily  clear.  However brown dwarf  kinematics provide important
insights into their formation, cooling, and disk heating.

The effect of  atmospheric cooling on populations of  brown dwarfs has
been discussed  in great detail  throughout the literature,  but there
are  three works  that are  particularly relevant  to our  study.  (1)
\citet{burg04} present a series of  Monte Carlo simulations to predict
the present-day mass  function (PDMF) of brown  dwarfs.  Using various
assumptions  for the  initial  mass function  (IMF),  birth rate,  and
cooling   models   \citep{burrows97,baraffe03},   they   confirm   the
relatively low number density of  mid-L dwarfs, even for shallow IMFs.
(2)  \citet{deacon06}  develop  a  similar calculation  to  model  the
present-day luminosity function  (PDLF), which have been  used to rule
out extreme (halo-type) birth rates \citep{dayjones13} and confirm the
low  space density  of L/T  transition dwarfs  \citep{marocco15}.  The
increasing  number of  late-T field  dwarfs has  been seen  in several
studies \citep[e.g.][]{metchev08,kirk12,burn13}, which  is seemingly a
{\it pile up} of cool  dwarfs.  (3) \citet{burg15} propose a two-phase
star-formation   history   to   explain   the   old   kinematic   ages
($6.5\pm0.4$~Gyr) of  L dwarfs  in their sample  of 85  M8--L6 dwarfs.
Therefore the aim of our work is to build on such kinematic models and
predict the  vertical number density  of brown dwarfs in  the Galactic
disk.

We organize this paper as  follows:~in \Sect{sec:sims} we describe our
Monte Carlo simulation, in \Sect{sec:results} we present our estimates
for the disk  thickness, in \Sect{sec:disc} we discuss  our results in
the  context of  what  can be  achieved with  the  next generation  of
surveys, and in \Sect{sec:summary} we summarize our key findings.

\section{Monte Carlo Simulations} \label{sec:sims}

To characterize the  effects of brown dwarf cooling  on their Galactic
properties, we develop a simple Monte Carlo simulation that is similar
to                several                 other                studies
\citep[e.g.][]{reid99,burg04,dayjones13,burg15}.  We  start  with  the
usual assumption that  the creation function is separable  in mass and
time:
\begin{equation}
C(m,t)\,\dd m\,\dd t = \phi(m)\,\psi(t)\,\dd m\,\dd t,
\end{equation}
where $\phi(m)$ is the IMF and $\psi(t)$ is the star-formation history
(SFH).  Although these  two distributions are often the  focus of many
studies,  we  only  wish  to  adopt  plausible  functional  forms  and
reasonable  range  of parameters  to  set  bounds  on the  effects  of
atmospheric cooling.  Therefore we use a powerlaw IMF
\begin{equation}\label{eqn:imf}
  \phi(m;\alpha)\,\dd m\propto m^{-\alpha}\, \dd m,
\end{equation}
for  $0.0005\!\leq\!m\!\leq\!0.1~M_{\odot}$,  and we  consider  models
with  $\alpha\!\in\![0,0.5,1.0,1.5]$, but  discuss the  role of  these
mass  limits  in \App{sec:masslimits}.   For  these  mass limits,  our
simulation is valid for spectral types $\sim$M8--T9.

We adopt a gamma-distribution form for the SFH:
\begin{equation}\label{eqn:sfh}
\psi(t;\beta,\tau)\,\dd t\propto\left(\frac{t}{\tau}\right)^{\beta}\exp\left(-\frac{t}{\tau}\right)\,\frac{\dd t}{\tau}
\end{equation}
for $0\!\leq\!t\!\leq\!t_0$, where $t_0$ is  the age of the Milky Way,
which we take to be 12~Gyr. To avoid confusion, we use $t$ to refer to
the time since the formation of the  Milky Way disk and $a$ as the age
of a brown dwarf. These  two quantities are related as: $a\!=\!t_0-t$.
We                 consider                models                 with
$\tau\!\in\![0.3,0.5,0.7,0.9,1.1]~\mathrm{Gyr}$                    and
$\beta\!\in\![2.5,3.5,4.5,5.5,6.5]$  (an exponential  model being  the
special  case of  $\beta\!=\!0$).  The  parameters $\beta$  and $\tau$
represent the power-law slope at  early times ($t\!\ll\!\tau$) and the
exponential decay at late times ($t\!\sim\!\tau$), respectively.  This
parameterization  of the  SFH  has a  maximum  star-formation rate  at
$t_{max}\!=\!\tau\beta$,  and  it  is qualitatively  similar  to  that
proposed by \citet{snaith}.  Also, this SFH is consistent with that of
the  typical Milky  Way-like galaxy  \citep{pap15}, which  we estimate
from            their             work            will            have
$(\beta,\tau)\!\approx\!(4.5,0.7~\mathrm{Gyr})$.

With these two distribution functions, the creation of simulated brown
dwarfs is parameterized by three tunable values $(\alpha,\beta,\tau)$,
which  we refer  to as  the  \textit{galaxy model}.   For each  galaxy
model, we  draw $10^{9}$ random  masses and formation times  (or ages)
from Equations~\ref{eqn:imf} and \ref{eqn:sfh}, respectively.

\begin{figure}
%  \epsscale{1.2}
%  \plotone{ccs_burrows97.ps}
  \includegraphics[width=0.46\textwidth]{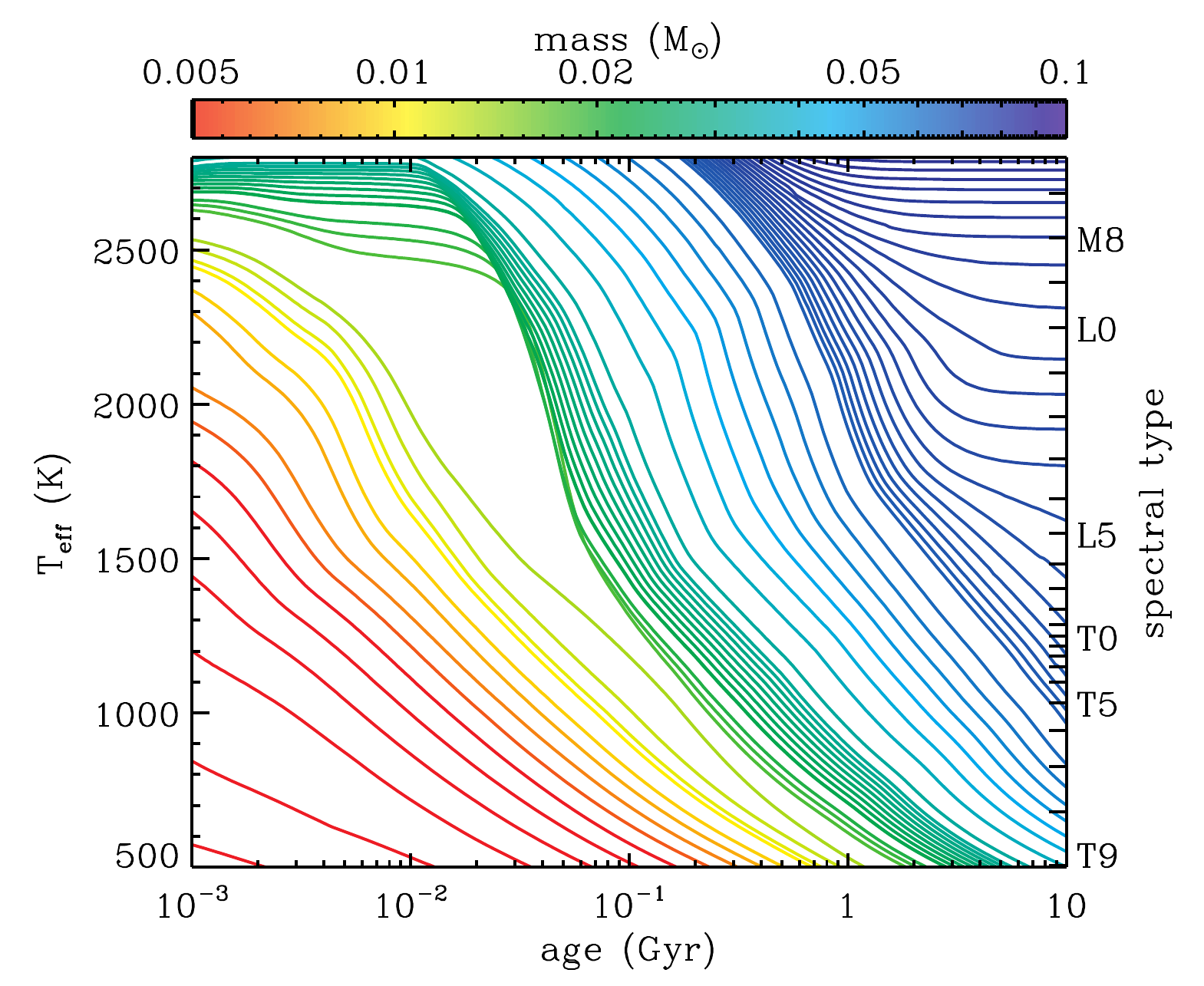}
  \caption{Cooling curves  from \citet{burrows97}.   Here we  show the
    effective temperature as  a function of age.   The lines represent
    the tracks expected for different  stellar masses, as indicated by
    their  color and  the  color map  at the  top.   The dashed  lines
    indicate the  min/max mass  for our  study.  We  transform between
    effective temperature and spectral type using the relation derived
    by \citet{filippazzo}.   For our  non-cooling models, we  hold the
    temperature   and   spectral   type   fixed  to   the   value   at
    1~Myr.\label{fig:ccs}}
\end{figure}

We adopt the cooling models of \citet{burrows97} for solar metallicity
to describe  the effective temperature  as a function of  stellar mass
and  age.   In  \Fig{fig:ccs},  we  show  these  cooling  models  with
appropriate spectral  type ranges indicated (see  \Eqn{eqn:teff} below
for additional details).  Although we expect metallicity to affect the
way  brown  dwarfs  cool,  our  primary  goal  is  to  illustrate  the
order-of-magnitude  effect  of  cooling and  will  discuss  additional
concerns with metallicity-dependent  models in Section~\ref{sec:disc}.
To  determine   the  current   effective  temperatures,   we  linearly
interpolate these cooling models for the simulated masses and ages.

To compare  our simulations to  observations, we must  convert between
effective  temperature ($T_{eff}$)  and spectral  type (\spt),  and we
adopt  the polynomial  model derived  by \citet{filippazzo}  for field
dwarfs:
\begin{eqnarray}\label{eqn:teff}
  T_{eff}(\spt) &=& 4747 - 700.5\times(\spt-60)\nonumber\\
  & & +115.5\times(\spt-60)^2\nonumber\\
  & & -11.91\times(\spt-60)^3\nonumber\\
  & & +0.6318\times(\spt-60)^4\nonumber\\
  & & -0.01606\times(\spt-60)^5\nonumber\\
  & & +0.0001546\times(\spt-60)^6,
\end{eqnarray}
where $T_{eff}$  is Kelvin  and $\spt$ is  the numeric  spectral type.
Throughout  this work,  we encode  spectral  type for  dwarf stars  as
starting at  0 (for O0V), incrementing  by 10 for each  major spectral
type  and by  1 for  each  minor type.   For  example, a  G2V star  is
represented  as  $\spt\!=\!42$.   The \citet{filippazzo}  relation  is
valid for $66\!\leq\!\spt\!\leq\!89$ (M6--T9).  We also considered the
\citet{stephens09} temperature-type relation  and find similar results
to those discussed below.

\begin{figure}
%  \epsscale{1.15}
%  \plotone{diflaw.ps}
  \includegraphics[width=0.46\textwidth]{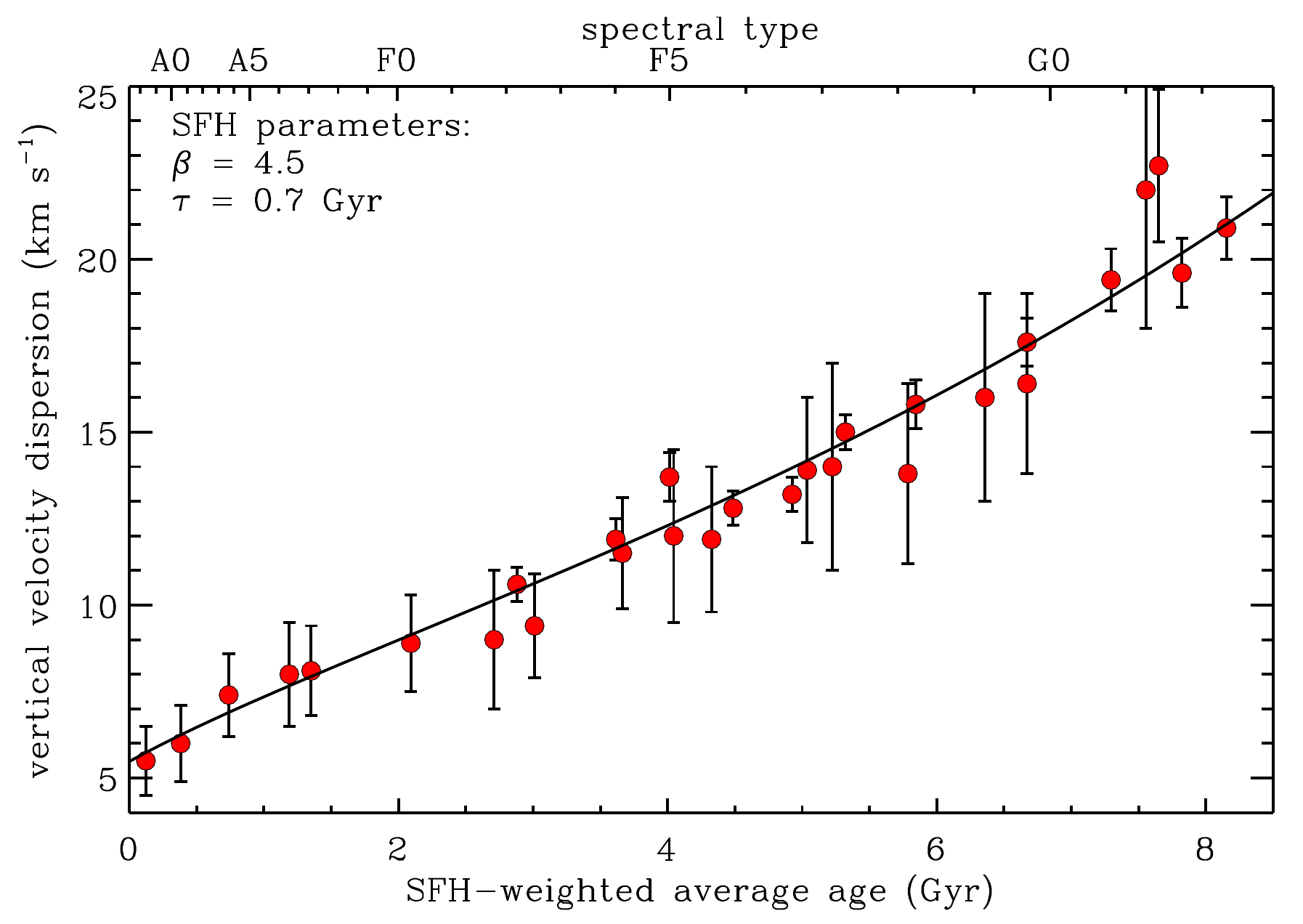}
  \caption{Calibrated diffusion  law.  The data points  here are taken
    from \citet{db98} and \citet{ab09},  and the solid line represents
    the best fit  to the velocity-dependent diffusion  model where the
    diffusion   constant    varies   with    time   \citep[Equation~16
      of][]{wiel77}.    We   convert   between   spectral   type   and
    SFH-weighted   average   age   using    the   assumed   SFH   (see
    \Eqn{eqn:avet}),  and   therefore  this  diffusion  law   must  be
    calibrated  for each  unique  combination of  the SFH  parameters.
    Here       we      show       the      nominal       model      of
    $(\beta,\tau)\!=\!(4.5,0.7~\mathrm{Gyr})$,  which   is  consistent
    with a Milky Way-like galaxy \citep{pap15}. \label{fig:diflaw}}
\end{figure}

We assign vertical velocities to our simulated stars using the results
of  \citet{wiel77},   who  solves   the  Fokker-Planck   equation  and
demonstrates that the vertical velocity dispersion increases with age.
Under different assumptions for the velocity and/or time dependence of
the    diffusion   coefficient,    \citet{wiel77}   derives    various
relationships between the vertical velocity  dispersion and the age of
the population.  We adopt the velocity-dependent diffusion coefficient
that varies with time, as it offers the highest degree of flexibility,
which gives a velocity dispersion of the form:
\begin{equation}\label{eqn:diffusion}
  \sigma_v^3 = \sigma_{v,0}^3+\frac{3}{2}\,\gamma_{v,p}\,T_{\gamma}\,\left(\exp\left(\frac{a}{T_{\gamma}}\right)-1\right),
\end{equation}
where ($\sigma_{v,0},\gamma_{v,p},T_{\gamma}$)  are tunable parameters
to be determined from observations.  Given the observations of cluster
brown dwarfs \citep[e.g.][]{mc05}, we  assume that brown dwarf initial
kinematics are  comparable to  main-sequence stars.  Therefore  we use
the vertical velocity dispersions  of \citet{db98} and \citet{ab09} as
a calibration set.  However, these  results are reported as a function
of spectral  type, not age  as we  require.  Therefore we  compute the
average stellar population birth time, weighted over the adopted SFH:
\begin{equation}\label{eqn:avet}
\left<t(\spt)\right> = \frac{\int_{t'}^{t_0} t\,\psi(t)\,\dd t}{\int_{t'}^{t_0}\psi(t)\,\dd t}
\end{equation}
where the lower limit  of integration is $t'\!=\!t_0-t_{MS}(\spt)$
and  $t_{MS}(\spt)$ is  the main-sequence  lifetime as  a function  of
spectral type:
\begin{eqnarray}\label{eqn:msl}
  \log t_{MS}(\spt) &=& 0.92 + 0.040\times(\spt-40)\nonumber\\
  & & -0.00051\times(\spt-40)^2\nonumber\\
  & & +0.000045\times(\spt-40)^3
\end{eqnarray}
where $t_{MS}(\spt)$  is in gigayears ---  \Eqn{eqn:msl} is polynomial
fit to tabulted  data \citep{lang}.  Now the  SFH-weighted average age
is:  $\left<a(\spt)\right>\!=\!t_0-\left<t(\spt)\right>$.  Since  this
transformation  between  spectral type  and  age  depends on  the  SFH
through \Eqn{eqn:avet},  it must be  computed for each  combination of
$(\beta,\tau)$.   In  \Tab{tab:res},  we   give  best-fit  values  for
$(\sigma_{w,0},\gamma_{w,p},T_{\gamma})$  and  reduced  $\chi^2_{\nu}$
determined      for      stars     with      $t_{MS}(\spt)\!\leq\!t_0$
($\spt\!\lesssim\!44$).     The    values   originally    quoted    by
\citet{wiel77} are  for the three-space velocity  dispersion, but here
we explicitly only deal with $w$-component of the velocity dispersion.
In fact, the vertical velocity dispersions quoted by \citet{wiel77} in
their Table~1 yield similar results to our estimates in \Tab{tab:res}:
$\sigma_{w,0}\!=\!9$~km~s$^{-1}$,
$\gamma_{w,p}\!=\!68$~(km~s$^{-1}$)$^3$~Gyr$^{-1}$,                and
$T_\gamma\!=\!2$~Gyr. In \Fig{fig:diflaw}, we  show the kinematic data
with    the    calibrated   diffusion    law    for    the   SFH    of
$(\beta,\tau)\!=\!(4.5,0.7~\mathrm{Gyr})$.

We  have verified  that  our  choice of  the  parameterization of  the
diffusion  law  does  not  dictate our  findings  discussed  below  by
considering the alternatives  \citep[e.g.][]{wiel77,ab09}.  Of course,
the  parameter values  of the  diffusion  law depend  strongly on  the
diffusion  model and  the SFH,  but  the key  aspect is  that we  must
capture the  trend of  increasing velocity dispersion  with population
age  or   main-sequence  lifetime  (see   Figures~\ref{fig:diflaw}  or
\ref{fig:example} for examples).

With the diffusion  model calibrated for the choice of  SFH, we have a
self-consistent description for the  velocity dispersion as a function
of stellar age. The velocity dispersion measured for a population over
a range of spectral types (or effective temperatures) will be:
\begin{equation}
   \left<\sigma^2_w\right>=\frac{1}{N}\sum_{i=1}^{N}\sigma^2_w(a_i)
\end{equation}
where $N$ is the number of samples in the desired spectral type range.
Now   the  simulated   parameter  $\left<\sigma^2_w\right>^{1/2}$   is
directly   comparable  to   published  estimates   for  the   velocity
dispersion.

We also consider a \textit{non-cooling}  model to establish a baseline
for  comparison.  Here  we repeat  the  above procedure,  but fix  the
effective temperature to the value  at 1~Myr and increase the velocity
dispersion according  to the same calibrated  \citet{wiel77} diffusion
laws.  At  this point we have  a library of simulated  dwarfs, each of
which  has been  assigned mass,  age, effective  temperature, spectral
type,  and vertical  velocity, all  of which  are consistent  with our
assumptions  on   the  IMF,   SFH,  cooling  models,   and  calibrated
observations (such as \Eqn{eqn:teff}  or \Eqn{eqn:msl}).  We construct
this library for the suite of  galaxy models listed above, and compute
various distributions to compare with observations.

As a final  note, we present the analytic  integral in \App{sec:integ}
that is approximated by this Monte Carlo simulation.

\section{Results}\label{sec:results}

Using our  libraries of simulated  disk brown  dwarfs, we are  able to
compute  several  quantities,  namely, the  distribution  of  spectral
types,  luminosities   \citep[assuming  the  $M_J(\mathrm{Vega})-\spt$
  relation  described  by][]{dupuy12},  and  PDLFs.   We  confirm  the
findings  presented by  \citet{burg04,dayjones13,marocco15}, but  omit
their presentation  here for brevity.   Instead, we turn to  develop a
unique  component stemming  from  the kinematics  of  the brown  dwarf
population.

\begin{figure}
%  \epsscale{1.1}
%  \plotone{vsig.ps}
  \includegraphics[width=0.46\textwidth]{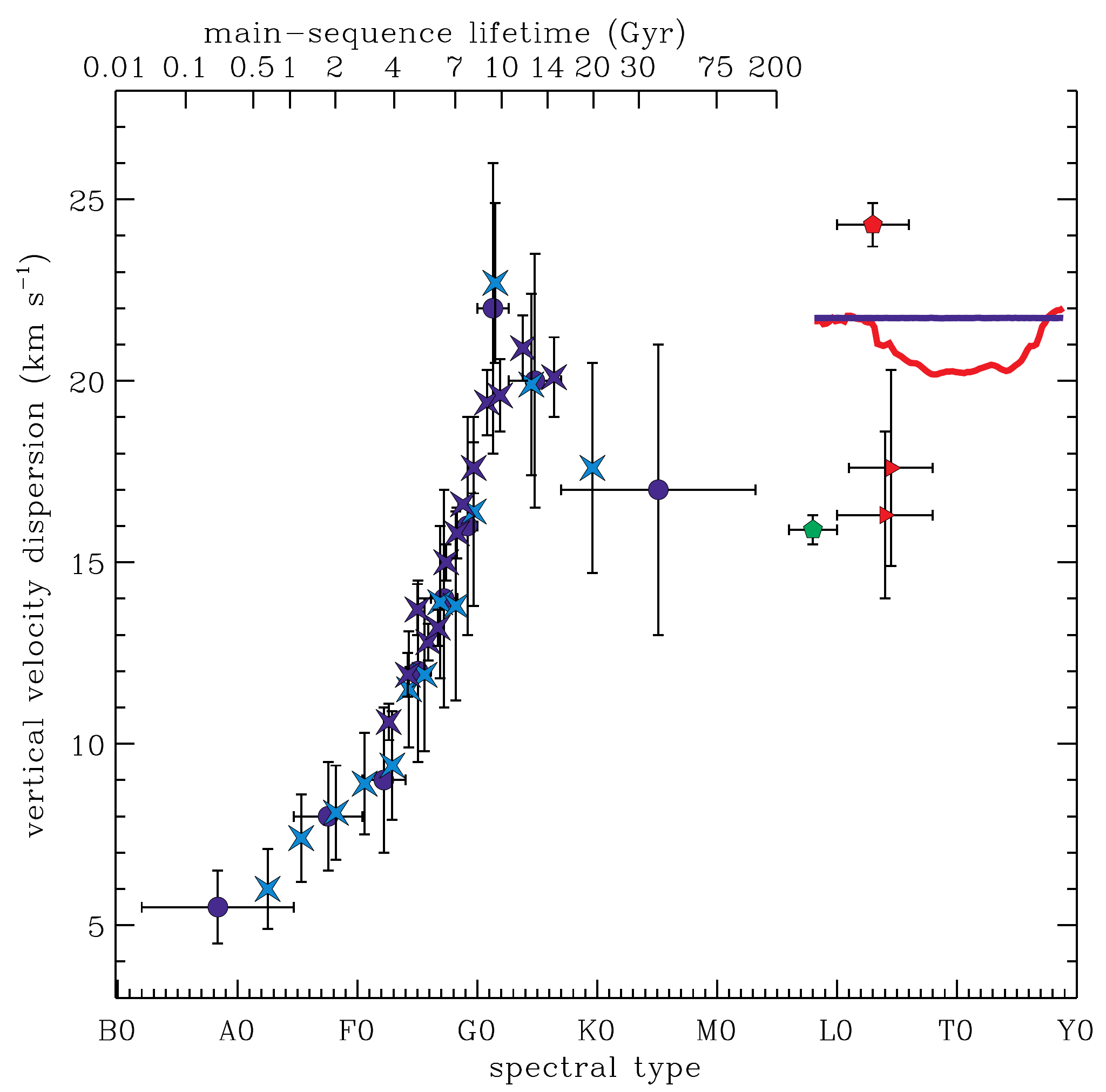}
  \caption{Example  vertical velocity  dispersion.  Here  we show  our
    model                        kinematics                        for
    $(\alpha,\beta,\tau)\!=\!(0.5,4.5,0.7~\mathrm{Gyr})$    with   the
    cooling (red  line) and non-cooling  (blue line) models  for brown
    dwarfs.    The  data   points  are   taken  from   the  literature
    (\citep[dark-blue       circles:][]{db98},       \citep[light-blue
      stars:][]{ab09},  \citep[right-facing  triangles:][]{rb09},  and
    \citep[pentagons:][]{burg15}).   We indicate  spectral types  that
    are above the hydrogen-burning limit as shades of blue, types that
    plausibly straddle the  limit as green, and  types likely entirely
    below the  limit as shades  of red.  The non-cooling  models (blue
    line)  are  always constant  and  the  cooling models  (red  line)
    deviate  significantly,  generally   reaching  a  minimum  between
    L5--T0.   It is  important  to  stress, our  intention  is not  to
    reproduce the brown dwarf  observations, but rather to demonstrate
    the  effects of  cooling on  their kinematics.   We consider  this
    deviation as  evidence that  the atmospheric  cooling may  leave a
    detectable      signature       in      the       brown      dwarf
    kinematics.  \label{fig:example}}
\end{figure}

In  \Fig{fig:example},   we  show   the  expected   vertical  velocity
dispersion as a  function of spectral type, where the  data points are
taken from the literature (see the figure caption for details) and the
red/blue lines indicate  our cooling/non-cooling models, respectively.
All of our  galaxy models show a few  characteristic behaviors: First,
the  velocity  dispersions  from  the non-cooling  models  are  always
constant, and roughly consistent with what is seen in low-mass dwarfs.
Second, the cooling models have a roughly constant velocity dispersion
for $\sim$M8--L2,  then decrease significantly. The  shape, depth, and
minimum spectral type of this deviation depend on the adopted galactic
model ($\alpha,\beta,\tau$).  Our goal here  is not to constrain these
model parameters and/or attempt to reproduce the brown dwarf kinematic
measurements, but rather point out  that this deviation occurs for any
reasonable choice of the galaxy model.  Additionally, the deviation is
only present  when we allow the  brown dwarfs to cool,  which suggests
that atmospheric  cooling may leave  an imprint on  the Galactic-scale
distribution  of brown  dwarfs  at fixed  spectral  type.  Third,  the
T/Y-transition objects generally  have velocity dispersions consistent
with the  M-dwarfs, suggesting  that atmospheric cooling  for L-dwarfs
proceeds    faster     than    diffusion    throughout     the    disk
\citep[e.g.][]{burg04}.

\begin{figure}
%  \epsscale{1.15}
%  \plotone{deviation.ps}
  \includegraphics[width=0.46\textwidth]{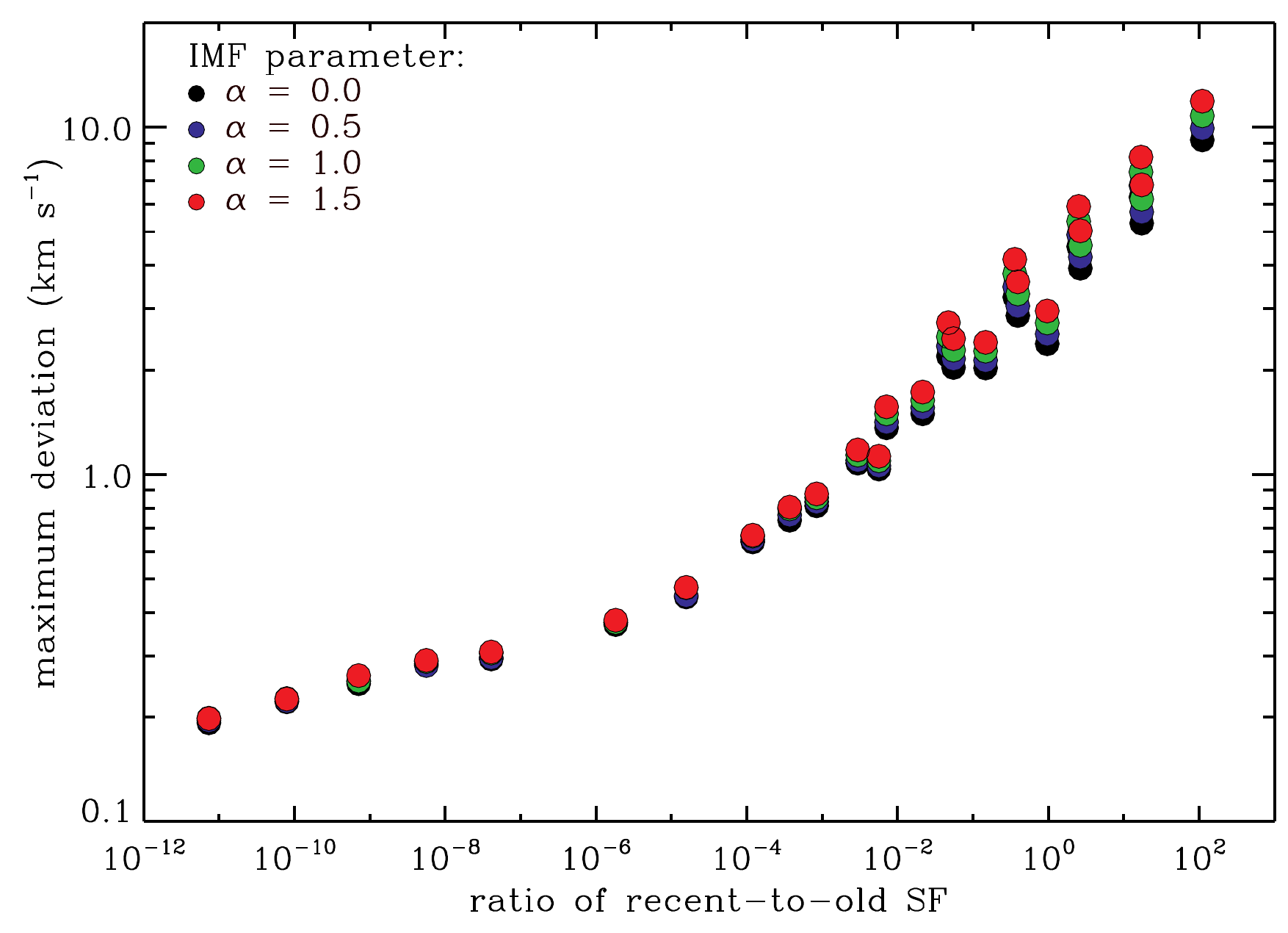}
  \caption{Velocity   dispersion  deviation.   We  show   the  maximum
    deviations  (defined   as  the  maximum  difference   between  the
    non-cooling  and cooling  models) as  a function  of the  ratio of
    recent-to-old star formation (as defined by \Eqn{eqn:ratio}).  The
    color of the plot symbol refers  to the power-law slope of the IMF
    (black:    $\alpha\!=\!0$,    blue:    $\alpha\!=\!0.5$,    green:
    $\alpha\!=\!1$,  and  red:  $\alpha\!=\!1.5$).  The  SFH  has  the
    largest  influence on  the properties  of the  velocity dispersion
    deviation  illustrated in  \Fig{fig:example}, with  the IMF  slope
    playing  a  lesser  role.   The additional  assumptions,  such  as
    diffusion law and mass limits, are largely irrelevant in dictating
    the velocity dispersion deviations.  \label{fig:deviation}}
\end{figure}

As  mentioned above,  the properties  of the  deviation depend  on the
choice of galaxy  model.  We find that the power-law  slope of the IMF
has the least importance, and our range of IMF slopes can only account
for  a  $\lesssim\!1\%$   change  in  the  depth   of  the  deviation.
Furthermore the  SFH parameters ($\beta,\tau$) have  considerably more
effect details of the deviation,  which is qualitatively consistent to
the \citet{burg15} finding.  To  investigate that somewhat further, we
consider the ratio of star-formation in  the last 2~Gyr to that in the
first 2~Gyr of the life of the Milky Way:
\begin{equation}\label{eqn:ratio}
  R=\frac{\int_{10}^{12}\,\psi(t)\,\dd t}{\int_{0}^{2}\,\psi(t)\,\dd t},
\end{equation}
as a  means of tracing the  fraction of young-to-old brown  dwarfs. In
\Fig{fig:deviation}  we show  the  maximum deviation  in the  velocity
dispersion (computed as the maximum difference between the non-cooling
and  cooling models)  as  a  function of  the  ratio of  recent-to-old
star-formation; the data points are color  coded by the IMF slope.  We
see  that the  deviation is  largest for  SFHs that  are dominated  by
recent  star  formation, further  suggesting  that  cooling is  a  key
effect.  We find that maximum  deviation generally occurs for spectral
types L5--T0.

These  results  are  almost  completely insensitive  to  many  of  our
underlying assumptions.  First, the range of  masses in the IMF has no
significant effect on these results  provided the mass range is within
the   range   reliably   sampled    by   the   cooling   models   (see
\App{sec:masslimits}).  Second, the  choice of the velocity-dispersion
parameterization has little effect, other  than to say it is necessary
that  the velocity  dispersion must  predict the  precipitous increase
seen in spectral types $\lesssim$G4V shown in \Fig{fig:example}.

\begin{figure}
%  \epsscale{1.1}
%  \plotone{zscl.ps}
  \includegraphics[width=0.46\textwidth]{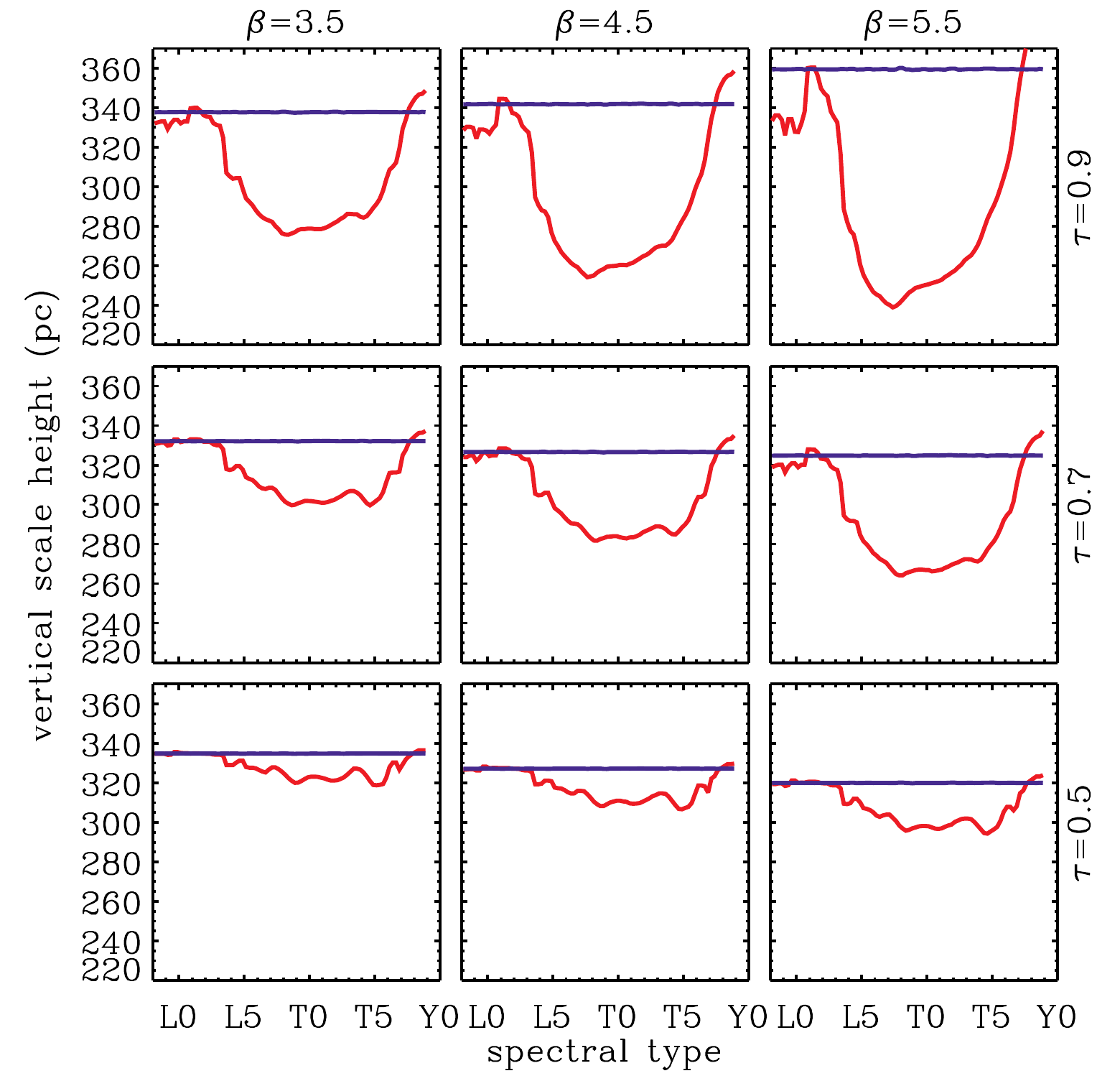}
  \caption{Predicted vertical  scale height. We convert  from velocity
    dispersion to scale height  assuming the $n\!=\!2$ model developed
    by   \citet{vdk88}   and   a   disk  surface   mass   density   of
    $\Sigma\!=\!68~M_{\odot}~\mathrm{pc}^{-2}$ \citep{bovy13}.   As in
    \Fig{fig:example}, the non-cooling/cooling models are indicated by
    solid blue/red lines, respectively.  Here we show $\alpha\!=\!0.5$
    with a  range of  ($\beta,\tau$), centered  on the  baseline value
    estimated by \citet{pap15}.  Current  modeling techniques have the
    accuracy to  detect a  change of  $\Delta z_{scl}\!\approx\!50$~pc
    \citep[e.g.][]{ryan11},  however few  surveys have  the sufficient
    sample  size  and  spectral  type  resolution/accuracy  needed  to
    resolve such deviations.\label{fig:zscl}}
\end{figure}

As outlined in  \Sect{sec:intro}, many studies find  that the vertical
velocity  dispersions of  L dwarfs  is larger  than that  of M  dwarfs
\citep[e.g.][]{seif09,blake10,burg15}.    To   examine  this   effect,
\citet{burg15} developed a very similar  Monte Carlo simulation to our
work,  and  they  see  hints  of the  deviation  described  above  for
comparable  Galactic models.   However  they propose  a two-phase  SFH
where brown dwarfs are  systematically older than main-sequence stars,
which they assume are formed at a constant rate throughout the life of
the Milky  Way.  In this  way, their two-phase Monte  Carlo simulation
predicts a roughly increasing  or constant velocity dispersion between
the late-M and  early-L dwarfs.  This effect seems  to naturally arise
in our models  (see \Fig{fig:example}), which is a  consequence of our
SFH  that has  more  late-time star  formation.   Moreover, our  model
predicts the strongest  deviation from constancy for  types later than
$\sim$L4, for which the \citet{burg15} sample is limited (6/85).

\begin{deluxetable*}{lccl}
\tablewidth{0pc}
\tablecaption{Estimates of the Vertical Scale Height$^{\dagger}$\label{tab:zscl}}
\tablehead{\colhead{Reference} & \colhead{spectral type} & \colhead{$z_{scl}$} & \colhead{Notes}\\
\colhead{} & \colhead{} & \colhead{(pc)} & \colhead{}}
\startdata
\citet{chen01} & F0--M4 & $330\pm3$ & type computed from \citet{covey07}\\ 
\citet{zheng01} & \nodata & 300 & spectral type not clear \\ 
\citet{ryan05} & $>$M6 & $350\pm50$ & \\
\citet{pirzkal05} & $>$M4 & $400\pm100$ & \\
\citet{juric08} & M0--M4 & $\sim\!250$ & additionally fit thick disk and halo, types computed \\
&&&from \citet{covey07} \\ 
\citet{pirzkal09} & M4--M9 & $370_{-65}^{+60}$ &  \\
\citet{pirzkal09} & M0--M9 & $300\pm70$ & \\ 
\citet{boch10} & M0--M8 & $300\pm15$ & additionally fit thick disk component\\
\citet{ryan11} & $>$M8 & $290\pm40$ &  \\
\citet{holw14} & M5--M9 & $400\pm100$ & use $\sech^2(\cdot)$ model \\
\citet{vv16} & M0--M9 & $290\pm20$ & additionally fit halo component\\
\enddata
\tablenotetext{\dagger}{Vertical scale height of thin disk component.}
\end{deluxetable*}

\section{Discussion}\label{sec:disc}

We have demonstrated that atmospheric  cooling may leave an imprint on
the kinematics as a function of  spectral type via disk heating, since
hot brown dwarfs must be  young.  By somewhat reversing this argument,
we  can use  observations  as a  function of  spectral  type to  place
constraints  on  the  spectral  type of  the  hydrogen-burning  limit,
averaged  over  metallicity.   Many  kinematic  surveys  do  not  have
sufficient sample sizes to  permit examination of velocity dispersions
in fine bins of spectral type, as  may be needed to see the deviations
predicted here.   Furthermore, most  kinematic samples are  limited to
the earliest L-dwarfs,  and we predict the deviations  to begin around
$\sim$L3  \citep[c.f.][]{burg15}.   However  it  may  be  possible  to
leverage the  known relationships  between disk  thickness ($z_{scl}$)
and  velocity dispersion,  then use  brown  dwarf number  counts as  a
similar   tool  \citep[e.g.][]{bahc86}.    \citet{vdk88}  proposes   a
flexible family of  vertical density distributions that  can explain a
host of observational phenomena:
\begin{equation}
  \rho(z)=2^{-2/n}\rho_{scl}\sech^{2/n}\left(\frac{n\,z}{2\,z_{scl}}\right),
\end{equation}
where  $n$  is  the  shape parameter.   The  classic  isothermal  disk
solution  and exponential  parameterization  are  special cases  where
$n\!=\!1$ and  $n\!=\!\infty$, respectively.  This class  of functions
leads to the following relationship:
\begin{equation}
  z_{scl}=\zeta_n\,\frac{{\sigma_{20}}^2}{\Sigma_{68}},
\end{equation}
where $\sigma_{20}$  and $\Sigma_{68}$ are the  velocity dispersion in
units  of 20~km~s$^{-1}$  (the average  value of  main-sequence stars:
$\mathrm{G0}\!\geq\!\spt\!\geq\!\mathrm{M9}$,  see  \Fig{fig:example})
and     the      surface     mass     density     in      units     of
68~$M_{\odot}~\mathrm{pc}^{-2}$  \citep{bovy13},   respectively.   The
normalization constant depends on the shape parameter:
\begin{equation}
  \zeta_n\approx\begin{cases}
  217~\mathrm{pc} & n=1\\
  277~\mathrm{pc} & n=2\\
  435~\mathrm{pc} & n=\infty.
  \end{cases}
\end{equation}

% normalization constants (for sigma=20km/s)
% n=1   -> 217 pc  
% n=2   -> 277 pc 
% n=inf -> 435 pc

% normalization constants (for sigma=21km/s)
% n=1   -> 239 pc  
% n=2   -> 305 pc  
% n=inf -> 480 pc

We convert our predicted velocity  dispersion to vertical scale height
using  the   $n\!=\!2$  model  \citep{vdk88,deg}   in  \Fig{fig:zscl}.
Assuming           the           baseline           model           of
$(\alpha,\beta,\tau)\!=\!(0.5,4.5,0.7~\mathrm{Gyr})$, we  predict that
a pure  late-L/early-T dwarf sample  should occupy  a disk that  has a
scale height  of $z_{scl}\!\sim\!280$~pc, which is  about $\sim\!15\%$
thinner  than what  is seen  for late-M  dwarfs (see  \Tab{tab:zscl}).
Additionally,  the   critical  point   in  \Fig{fig:zscl}   where  the
cooling-model (red  line) scale  heights deviate from  the non-cooling
model  (blue  line) is  $\sim$L3,  which  would  be indicative  of  an
absolute  lower-bound   on  the   hydrogen-burning  limit   for  solar
metallicity.  Through careful calibration of the dwarfs that are known
to  be burning  hydrogen (such  as  by the  lithium test),  it may  be
possible   to  relate   the  spectral   type  of   deviation  to   the
hydrogen-burning  limit.   Since  the   velocity  dispersions  of  T/Y
transition  objects is  similar to  that  of the  M-dwarfs, we  expect
Y-dwarfs will occupy  a disk similar to  the M-dwarfs ($\sim\!300$~pc,
see  \Tab{tab:zscl}).   In  \Tab{tab:res},   we  report  the  vertical
velocity dispersion and scale height for M8--L2 and L6--T0 types.

\begin{deluxetable*}{cccccccccccc}
\tablewidth{0pc}
\tablecaption{Results for Monte Carlo Simulations\label{tab:res}}
\tablehead{\colhead{} & \colhead{} & \colhead{} & \colhead{} & \colhead{} & \colhead{} & \colhead{} & \colhead{} & \multicolumn{2}{c}{M8--L2} & \multicolumn{2}{c}{L6--T0}\\
          \colhead{$\alpha$} & \colhead{$\beta$} & \colhead{$\tau$} & \colhead{$\sigma_{w,0}$} & \colhead{$\gamma_{w,p}$} & \colhead{$T_{\gamma}$} & \colhead{$\chi^2_{\nu}$} & \colhead{$\log R$} & \colhead{$\sigma_w$} & \colhead{$z_{scl}$} & \colhead{$\sigma_w$} & \colhead{$z_{scl}$}\\
           \colhead{} & \colhead{} & \colhead{(Gyr)} & \colhead{(km/s)} & \colhead{((km/s)$^3$/Gyr)} & \colhead{(Gyr)} & \colhead{} & \colhead{(km/s)} & \colhead{(km/s)} & \colhead{(pc)} & \colhead{(km/s)} & \colhead{(pc)}}
\startdata
0.5 & 2.5 & 0.3 & $4.9\pm0.9$ & $138\pm17$ & $4.2\pm0.4$& 0.70 & $-11.1$ & 22.5 & 349 & 22.4 & 347\\
0.5 & 2.5 & 0.5 & $5.1\pm0.9$ & $144\pm19$ & $3.9\pm0.4$& 0.62 & $ -5.7$ & 22.3 & 342 & 22.0 & 335\\
0.5 & 2.5 & 0.7 & $5.1\pm0.9$ & $152\pm20$ & $3.7\pm0.4$& 0.56 & $ -3.4$ & 22.2 & 339 & 21.6 & 321\\
0.5 & 2.5 & 0.9 & $5.2\pm0.9$ & $160\pm22$ & $3.4\pm0.4$& 0.53 & $ -2.1$ & 22.2 & 339 & 20.9 & 303\\
0.5 & 2.5 & 1.1 & $5.2\pm0.9$ & $166\pm23$ & $3.1\pm0.3$& 0.51 & $ -1.3$ & 22.3 & 342 & 20.2 & 283\\
0.5 & 3.5 & 0.3 & $5.0\pm0.9$ & $133\pm17$ & $4.0\pm0.4$& 0.65 & $-10.1$ & 22.3 & 343 & 22.2 & 340\\
0.5 & 3.5 & 0.5 & $5.2\pm0.8$ & $136\pm18$ & $3.6\pm0.4$& 0.55 & $ -4.8$ & 22.0 & 334 & 21.6 & 323\\
0.5 & 3.5 & 0.7 & $5.3\pm0.8$ & $141\pm20$ & $3.3\pm0.3$& 0.51 & $ -2.5$ & 21.9 & 331 & 21.0 & 303\\
0.5 & 3.5 & 0.9 & $5.4\pm0.8$ & $145\pm21$ & $2.9\pm0.3$& 0.49 & $ -1.3$ & 22.0 & 333 & 20.1 & 279\\
0.5 & 3.5 & 1.1 & $5.4\pm0.8$ & $145\pm22$ & $2.5\pm0.3$& 0.49 & $ -0.4$ & 22.2 & 340 & 19.3 & 257\\
0.5 & 4.5 & 0.3 & $5.1\pm0.9$ & $128\pm17$ & $3.8\pm0.4$& 0.60 & $ -9.1$ & 22.1 & 337 & 22.0 & 333\\
0.5 & 4.5 & 0.5 & $5.3\pm0.8$ & $127\pm18$ & $3.3\pm0.3$& 0.51 & $ -3.9$ & 21.7 & 327 & 21.2 & 311\\
0.5 & 4.5 & 0.7 & $5.5\pm0.8$ & $129\pm20$ & $2.8\pm0.3$& 0.49 & $ -1.7$ & 21.7 & 325 & 20.3 & 285\\
0.5 & 4.5 & 0.9 & $5.6\pm0.8$ & $128\pm21$ & $2.4\pm0.2$& 0.50 & $ -0.4$ & 21.9 & 332 & 19.3 & 257\\
0.5 & 4.5 & 1.1 & $5.7\pm0.8$ & $121\pm21$ & $1.9\pm0.2$& 0.54 & $ +0.4$ & 22.5 & 350 & 18.8 & 243\\
0.5 & 5.5 & 0.3 & $5.2\pm0.9$ & $122\pm17$ & $3.6\pm0.4$& 0.55 & $ -8.2$ & 21.9 & 331 & 21.7 & 326\\
0.5 & 5.5 & 0.5 & $5.5\pm0.8$ & $118\pm18$ & $3.0\pm0.3$& 0.48 & $ -3.1$ & 21.5 & 319 & 20.8 & 299\\
0.5 & 5.5 & 0.7 & $5.7\pm0.8$ & $116\pm19$ & $2.5\pm0.2$& 0.51 & $ -0.8$ & 21.5 & 321 & 19.7 & 267\\
0.5 & 5.5 & 0.9 & $5.8\pm0.7$ & $108\pm20$ & $1.9\pm0.2$& 0.59 & $ +0.4$ & 22.1 & 338 & 18.8 & 243\\
0.5 & 5.5 & 1.1 & $6.0\pm0.7$ & $ 93\pm19$ & $1.4\pm0.1$& 0.69 & $ +1.2$ & 23.6 & 386 & 19.0 & 249\\
0.5 & 6.5 & 0.3 & $5.3\pm0.8$ & $117\pm17$ & $3.4\pm0.3$& 0.52 & $ -7.4$ & 21.7 & 325 & 21.5 & 319\\
0.5 & 6.5 & 0.5 & $5.6\pm0.8$ & $109\pm18$ & $2.7\pm0.3$& 0.49 & $ -2.3$ & 21.3 & 313 & 20.4 & 286\\
0.5 & 6.5 & 0.7 & $5.9\pm0.7$ & $101\pm19$ & $2.1\pm0.2$& 0.59 & $ -0.0$ & 21.6 & 321 & 19.1 & 251\\
0.5 & 6.5 & 0.9 & $6.1\pm0.7$ & $ 86\pm18$ & $1.5\pm0.1$& 0.75 & $ +1.2$ & 22.9 & 362 & 18.8 & 243\\
0.5 & 6.5 & 1.1 & $6.3\pm0.7$ & $ 64\pm16$ & $1.1\pm0.1$& 0.95 & $ +2.0$ & 26.7 & 492 & 20.7 & 296\\
1.0 & 2.5 & 0.3 & $4.9\pm0.9$ & $138\pm17$ & $4.2\pm0.4$& 0.70 & $-11.1$ & 22.5 & 349 & 22.4 & 347\\
1.0 & 2.5 & 0.5 & $5.1\pm0.9$ & $144\pm19$ & $3.9\pm0.4$& 0.62 & $ -5.7$ & 22.3 & 342 & 22.0 & 334\\
1.0 & 2.5 & 0.7 & $5.1\pm0.9$ & $152\pm20$ & $3.7\pm0.4$& 0.56 & $ -3.4$ & 22.2 & 339 & 21.5 & 320\\
1.0 & 2.5 & 0.9 & $5.2\pm0.9$ & $160\pm22$ & $3.4\pm0.4$& 0.53 & $ -2.1$ & 22.2 & 339 & 20.9 & 301\\
1.0 & 2.5 & 1.1 & $5.2\pm0.9$ & $166\pm23$ & $3.1\pm0.3$& 0.51 & $ -1.3$ & 22.2 & 341 & 20.1 & 278\\
1.0 & 3.5 & 0.3 & $5.0\pm0.9$ & $133\pm17$ & $4.0\pm0.4$& 0.65 & $-10.1$ & 22.3 & 343 & 22.2 & 340\\
1.0 & 3.5 & 0.5 & $5.2\pm0.8$ & $136\pm18$ & $3.6\pm0.4$& 0.55 & $ -4.8$ & 22.0 & 334 & 21.6 & 323\\
1.0 & 3.5 & 0.7 & $5.3\pm0.8$ & $141\pm20$ & $3.3\pm0.3$& 0.51 & $ -2.5$ & 21.9 & 331 & 20.9 & 302\\
1.0 & 3.5 & 0.9 & $5.4\pm0.8$ & $145\pm21$ & $2.9\pm0.3$& 0.49 & $ -1.3$ & 22.0 & 333 & 20.0 & 275\\
1.0 & 3.5 & 1.1 & $5.4\pm0.8$ & $145\pm22$ & $2.5\pm0.3$& 0.49 & $ -0.4$ & 22.1 & 338 & 19.0 & 249\\
1.0 & 4.5 & 0.3 & $5.1\pm0.9$ & $128\pm17$ & $3.8\pm0.4$& 0.60 & $ -9.1$ & 22.1 & 337 & 22.0 & 333\\
1.0 & 4.5 & 0.5 & $5.3\pm0.8$ & $127\pm18$ & $3.3\pm0.3$& 0.51 & $ -3.9$ & 21.7 & 327 & 21.2 & 311\\
1.0 & 4.5 & 0.7 & $5.5\pm0.8$ & $129\pm20$ & $2.8\pm0.3$& 0.49 & $ -1.7$ & 21.7 & 324 & 20.2 & 283\\
1.0 & 4.5 & 0.9 & $5.6\pm0.8$ & $128\pm21$ & $2.4\pm0.2$& 0.50 & $ -0.4$ & 21.9 & 330 & 19.1 & 251\\
1.0 & 4.5 & 1.1 & $5.7\pm0.8$ & $121\pm21$ & $1.9\pm0.2$& 0.54 & $ +0.4$ & 22.3 & 345 & 18.3 & 231\\
1.0 & 5.5 & 0.3 & $5.2\pm0.9$ & $122\pm17$ & $3.6\pm0.4$& 0.55 & $ -8.2$ & 21.9 & 331 & 21.7 & 326\\
1.0 & 5.5 & 0.5 & $5.5\pm0.8$ & $118\pm18$ & $3.0\pm0.3$& 0.48 & $ -3.1$ & 21.5 & 319 & 20.8 & 298\\
1.0 & 5.5 & 0.7 & $5.7\pm0.8$ & $116\pm19$ & $2.5\pm0.2$& 0.51 & $ -0.8$ & 21.5 & 320 & 19.5 & 263\\
1.0 & 5.5 & 0.9 & $5.8\pm0.7$ & $108\pm20$ & $1.9\pm0.2$& 0.59 & $ +0.4$ & 22.0 & 335 & 18.4 & 234\\
1.0 & 5.5 & 1.1 & $6.0\pm0.7$ & $ 93\pm19$ & $1.4\pm0.1$& 0.69 & $ +1.2$ & 23.3 & 376 & 18.4 & 233\\
1.0 & 6.5 & 0.3 & $5.3\pm0.8$ & $117\pm17$ & $3.4\pm0.3$& 0.52 & $ -7.4$ & 21.7 & 325 & 21.5 & 319\\
1.0 & 6.5 & 0.5 & $5.6\pm0.8$ & $109\pm18$ & $2.7\pm0.3$& 0.49 & $ -2.3$ & 21.3 & 313 & 20.3 & 285\\
1.0 & 6.5 & 0.7 & $5.9\pm0.7$ & $101\pm19$ & $2.1\pm0.2$& 0.59 & $ -0.0$ & 21.5 & 320 & 18.9 & 246\\
1.0 & 6.5 & 0.9 & $6.1\pm0.7$ & $ 86\pm18$ & $1.5\pm0.1$& 0.75 & $ +1.2$ & 22.7 & 356 & 18.3 & 230\\
1.0 & 6.5 & 1.1 & $6.3\pm0.7$ & $ 64\pm16$ & $1.1\pm0.1$& 0.95 & $ +2.0$ & 26.1 & 472 & 19.8 & 270\\
1.5 & 2.5 & 0.3 & $4.9\pm0.9$ & $138\pm17$ & $4.2\pm0.4$& 0.70 & $-11.1$ & 22.5 & 349 & 22.4 & 347\\
1.5 & 2.5 & 0.5 & $5.1\pm0.9$ & $144\pm19$ & $3.9\pm0.4$& 0.62 & $ -5.7$ & 22.3 & 342 & 22.0 & 334\\
1.5 & 2.5 & 0.7 & $5.1\pm0.9$ & $152\pm20$ & $3.7\pm0.4$& 0.56 & $ -3.4$ & 22.2 & 339 & 21.5 & 320\\
1.5 & 2.5 & 0.9 & $5.2\pm0.9$ & $160\pm22$ & $3.4\pm0.4$& 0.53 & $ -2.1$ & 22.2 & 339 & 20.8 & 299\\
1.5 & 2.5 & 1.1 & $5.2\pm0.9$ & $166\pm23$ & $3.1\pm0.3$& 0.51 & $ -1.3$ & 22.2 & 340 & 19.9 & 272\\
1.5 & 3.5 & 0.3 & $5.0\pm0.9$ & $133\pm17$ & $4.0\pm0.4$& 0.65 & $-10.1$ & 22.3 & 343 & 22.2 & 340\\
1.5 & 3.5 & 0.5 & $5.2\pm0.8$ & $136\pm18$ & $3.6\pm0.4$& 0.55 & $ -4.8$ & 22.0 & 334 & 21.6 & 323\\
1.5 & 3.5 & 0.7 & $5.3\pm0.8$ & $141\pm20$ & $3.3\pm0.3$& 0.51 & $ -2.5$ & 21.9 & 331 & 20.9 & 301\\
1.5 & 3.5 & 0.9 & $5.4\pm0.8$ & $145\pm21$ & $2.9\pm0.3$& 0.49 & $ -1.3$ & 21.9 & 332 & 19.8 & 271\\
1.5 & 3.5 & 1.1 & $5.4\pm0.8$ & $145\pm22$ & $2.5\pm0.3$& 0.49 & $ -0.4$ & 22.0 & 335 & 18.6 & 240\\
1.5 & 4.5 & 0.3 & $5.1\pm0.9$ & $128\pm17$ & $3.8\pm0.4$& 0.60 & $ -9.1$ & 22.1 & 337 & 22.0 & 333\\
1.5 & 4.5 & 0.5 & $5.3\pm0.8$ & $127\pm18$ & $3.3\pm0.3$& 0.51 & $ -3.9$ & 21.7 & 327 & 21.2 & 311\\
1.5 & 4.5 & 0.7 & $5.5\pm0.8$ & $129\pm20$ & $2.8\pm0.3$& 0.49 & $ -1.7$ & 21.7 & 324 & 20.2 & 281\\
1.5 & 4.5 & 0.9 & $5.6\pm0.8$ & $128\pm21$ & $2.4\pm0.2$& 0.50 & $ -0.4$ & 21.8 & 328 & 18.8 & 243\\
1.5 & 4.5 & 1.1 & $5.7\pm0.8$ & $121\pm21$ & $1.9\pm0.2$& 0.54 & $ +0.4$ & 22.1 & 338 & 17.7 & 217\\
1.5 & 5.5 & 0.3 & $5.2\pm0.9$ & $122\pm17$ & $3.6\pm0.4$& 0.55 & $ -8.2$ & 21.9 & 331 & 21.7 & 326\\
1.5 & 5.5 & 0.5 & $5.5\pm0.8$ & $118\pm18$ & $3.0\pm0.3$& 0.48 & $ -3.1$ & 21.5 & 319 & 20.8 & 298\\
1.5 & 5.5 & 0.7 & $5.7\pm0.8$ & $116\pm19$ & $2.5\pm0.2$& 0.51 & $ -0.8$ & 21.5 & 320 & 19.4 & 260\\
1.5 & 5.5 & 0.9 & $5.8\pm0.7$ & $108\pm20$ & $1.9\pm0.2$& 0.59 & $ +0.4$ & 21.9 & 330 & 18.0 & 223\\
1.5 & 5.5 & 1.1 & $6.0\pm0.7$ & $ 93\pm19$ & $1.4\pm0.1$& 0.69 & $ +1.2$ & 22.9 & 362 & 17.6 & 213\\
1.5 & 6.5 & 0.3 & $5.3\pm0.8$ & $117\pm17$ & $3.4\pm0.3$& 0.52 & $ -7.4$ & 21.7 & 325 & 21.5 & 318\\
1.5 & 6.5 & 0.5 & $5.6\pm0.8$ & $109\pm18$ & $2.7\pm0.3$& 0.49 & $ -2.3$ & 21.3 & 313 & 20.3 & 285\\
1.5 & 6.5 & 0.7 & $5.9\pm0.7$ & $101\pm19$ & $2.1\pm0.2$& 0.59 & $ -0.0$ & 21.5 & 318 & 18.6 & 240\\
1.5 & 6.5 & 0.9 & $6.1\pm0.7$ & $ 86\pm18$ & $1.5\pm0.1$& 0.75 & $ +1.2$ & 22.4 & 346 & 17.6 & 214\\
1.5 & 6.5 & 1.1 & $6.3\pm0.7$ & $ 64\pm16$ & $1.1\pm0.1$& 0.95 & $ +2.0$ & 25.4 & 444 & 18.6 & 240\\
0.0 & 2.5 & 0.3 & $4.9\pm0.9$ & $138\pm17$ & $4.2\pm0.4$& 0.70 & $-11.1$ & 22.5 & 349 & 22.4 & 347\\
0.0 & 2.5 & 0.5 & $5.1\pm0.9$ & $144\pm19$ & $3.9\pm0.4$& 0.62 & $ -5.7$ & 22.3 & 342 & 22.0 & 335\\
0.0 & 2.5 & 0.7 & $5.1\pm0.9$ & $152\pm20$ & $3.7\pm0.4$& 0.56 & $ -3.4$ & 22.2 & 339 & 21.6 & 321\\
0.0 & 2.5 & 0.9 & $5.2\pm0.9$ & $160\pm22$ & $3.4\pm0.4$& 0.53 & $ -2.1$ & 22.2 & 339 & 21.0 & 304\\
0.0 & 2.5 & 1.1 & $5.2\pm0.9$ & $166\pm23$ & $3.1\pm0.3$& 0.51 & $ -1.3$ & 22.3 & 343 & 20.4 & 286\\
0.0 & 3.5 & 0.3 & $5.0\pm0.9$ & $133\pm17$ & $4.0\pm0.4$& 0.65 & $-10.1$ & 22.3 & 343 & 22.2 & 340\\
0.0 & 3.5 & 0.5 & $5.2\pm0.8$ & $136\pm18$ & $3.6\pm0.4$& 0.55 & $ -4.8$ & 22.0 & 334 & 21.6 & 323\\
0.0 & 3.5 & 0.7 & $5.3\pm0.8$ & $141\pm20$ & $3.3\pm0.3$& 0.51 & $ -2.5$ & 21.9 & 331 & 21.0 & 304\\
0.0 & 3.5 & 0.9 & $5.4\pm0.8$ & $145\pm21$ & $2.9\pm0.3$& 0.49 & $ -1.3$ & 22.0 & 334 & 20.2 & 282\\
0.0 & 3.5 & 1.1 & $5.4\pm0.8$ & $145\pm22$ & $2.5\pm0.3$& 0.49 & $ -0.4$ & 22.3 & 342 & 19.5 & 263\\
0.0 & 4.5 & 0.3 & $5.1\pm0.9$ & $128\pm17$ & $3.8\pm0.4$& 0.60 & $ -9.1$ & 22.1 & 337 & 22.0 & 333\\
0.0 & 4.5 & 0.5 & $5.3\pm0.8$ & $127\pm18$ & $3.3\pm0.3$& 0.51 & $ -3.9$ & 21.7 & 327 & 21.2 & 312\\
0.0 & 4.5 & 0.7 & $5.5\pm0.8$ & $129\pm20$ & $2.8\pm0.3$& 0.49 & $ -1.7$ & 21.7 & 325 & 20.4 & 286\\
0.0 & 4.5 & 0.9 & $5.6\pm0.8$ & $128\pm21$ & $2.4\pm0.2$& 0.50 & $ -0.4$ & 22.0 & 333 & 19.5 & 262\\
0.0 & 4.5 & 1.1 & $5.7\pm0.8$ & $121\pm21$ & $1.9\pm0.2$& 0.54 & $ +0.4$ & 22.6 & 353 & 19.1 & 253\\
0.0 & 5.5 & 0.3 & $5.2\pm0.9$ & $122\pm17$ & $3.6\pm0.4$& 0.55 & $ -8.2$ & 21.9 & 331 & 21.7 & 326\\
0.0 & 5.5 & 0.5 & $5.5\pm0.8$ & $118\pm18$ & $3.0\pm0.3$& 0.48 & $ -3.1$ & 21.5 & 319 & 20.8 & 299\\
0.0 & 5.5 & 0.7 & $5.7\pm0.8$ & $116\pm19$ & $2.5\pm0.2$& 0.51 & $ -0.8$ & 21.6 & 321 & 19.8 & 269\\
0.0 & 5.5 & 0.9 & $5.8\pm0.7$ & $108\pm20$ & $1.9\pm0.2$& 0.59 & $ +0.4$ & 22.2 & 341 & 19.1 & 251\\
0.0 & 5.5 & 1.1 & $6.0\pm0.7$ & $ 93\pm19$ & $1.4\pm0.1$& 0.69 & $ +1.2$ & 23.9 & 394 & 19.6 & 264\\
0.0 & 6.5 & 0.3 & $5.3\pm0.8$ & $117\pm17$ & $3.4\pm0.3$& 0.52 & $ -7.4$ & 21.7 & 325 & 21.5 & 319\\
0.0 & 6.5 & 0.5 & $5.6\pm0.8$ & $109\pm18$ & $2.7\pm0.3$& 0.49 & $ -2.3$ & 21.3 & 313 & 20.4 & 287\\
0.0 & 6.5 & 0.7 & $5.9\pm0.7$ & $101\pm19$ & $2.1\pm0.2$& 0.59 & $ -0.0$ & 21.6 & 322 & 19.2 & 256\\
0.0 & 6.5 & 0.9 & $6.1\pm0.7$ & $ 86\pm18$ & $1.5\pm0.1$& 0.75 & $ +1.2$ & 23.1 & 367 & 19.2 & 255\\
0.0 & 6.5 & 1.1 & $6.3\pm0.7$ & $ 64\pm16$ & $1.1\pm0.1$& 0.95 & $ +2.0$ & 27.1 & 508 & 21.5 & 319\\
\enddata
\end{deluxetable*}

Again  there  is insufficient  spectral  type  resolution to  see  any
deviations on the order of a  few subtypes, which is due in-large part
to small-number statistics in the  small fields observed with the {\it
  Hubble  Space Telescope}  (HST).  For  example, \citet{ryan16}  show
that one  may expect  a surface density  of $\sim\!0.05$~arcmin$^{-2}$
per  five subtypes  (for the  lowest  density field),  or $1-2$  brown
dwarfs ($\spt\!\gtrsim$M8)  for a  typical high-latitude  field.  With
the  planned   {\it  Large-Scale  Synoptic  Telescope}   (LSST),  {\it
  Wide-Field Infrared Survey Telescope}  (WFIRST), or Euclid missions,
sufficient numbers and  type accuracy may make it  possible to observe
deviations in the  scale height as a function of  spectral type due to
atmospheric  cooling.   Indeed,  the infrared  spectroscopy  from  the
High-Latitude Survey (HLS) with WFIRST \citep{wfirst} may even provide
crude metallicity discrimination that would refine the cooling models.
The HLS  is expected  to survey $\sim\!2\,000$~deg$^2$,  therefore the
\citet{ryan16}  models  predict   $\sim\!35\,000$  L0--L5  dwarfs  and
$\sim\!30\,000$ for L5--L9 dwarfs.  The typical HST-based scale height
estimates  are  based   on  10s  of  dwarfs  and   an  uncertainty  of
$\sim\!50$~pc (see  \Tab{tab:zscl}).  If the uncertainty  on the scale
heights  scales $\propto\!1/\sqrt{N}$,  then  the typical  uncertainty
ought to  decrease to  $\sim\!5$~pc, and  so estimating  deviations on
this order is plausible. Conversely, \citet{best} recently published a
sample of  $\approx\!10\,000$ MLT-dwarfs with proper  motions measured
in Pan-STARRS, therefore  the sample broken by subtype  is expected to
be  considerably smaller.   Additionally,  slitless spectroscopy  from
{\it  James  Webb Space  Telescope}  (JWST)  will be  instrumental  in
constraining the faint-end of the  number counts, which would sample a
thick  disk  or Galactic  halo.   Nevertheless  it is  interesting  to
speculate that the macroscopic  (Galactic-scale) distribution of brown
dwarfs holds clues to the microscopic (brown dwarf-scale) physics.

As a final point, although the number counts and disk scale height may
probe  similar Galactic  physics as  the kinematic  measurements, they
will be  subject to entirely  different biases and  uncertainties. For
example,  three-dimensional velocity  estimates  are  often biased  to
samples of nearby objects, where  the total number of testable objects
may be limited and/or pathological in some way \citep[e.g.][]{seif09}.
In  fact,  the  very  broad range  of  reported  velocity  dispersions
\citep[$\sigma_w\!\sim\!15-25$~km~s$^{-1}$][]{zo07,faherty09,seif09,sjs10,blake10,burg15}
hints  at possiblity  for various  sample  biases, which  could be  as
benign  as small  number  statistics.   \citet{burg15} consider  three
additional  biases   in  their  kinematic  sample:   (1)  ``pointing''
asymmetries resulting  from source  confusion near the  Galactic plane
and/or  declination restrictions  from  the  observatory; (2)  ``youth
biases'' between the M and L  samples; and (3) a cosmic bias resulting
from  simply observing  multiple distinct  populations of  M/L dwarfs.
Ultimately, they  are unable to draw  any firm conclusions due  to the
small number of L dwarfs  (28/85).  However, such biases and/or sample
sizes are  likely explanations for  the tension between  our predicted
velocity dispersions and those often measured.

On the other hand, the brown  dwarf number counts are expected to peak
at  $J\!\simeq\!24$~AB   mag  \citep{ryan16},   which  is   a  readily
achievable   flux  limit   for   most  surveys   (whether  space-   or
ground-based).   Such   surveys  will   probe  distances   to  several
kiloparsecs and alleviate any  concerns for sample pathology, although
other forms of  biases may be introduced.  The most  likely concern is
some sort  of {\it  identification degeneracy}, where  observed colors
and  image  morphologies  do  not  exclusively  map  to  brown  dwarfs
\citep[e.g.][]{yan03,cab08}.  It  is hard to quantify  this degeneracy
further as  it depends critically  on the observational  parameters of
the survey, but  this is an issue that  slitless infrared spectroscopy
would remedy (such as with WFIRST).  Indeed spectroscopic observations
with HST/Wide-Field Camera  3 are already quite  successful at probing
distant and/or very cool dwarfs \citep[e.g.][]{masters}.

\section{Summary}\label{sec:summary}

We have presented a Monte  Carlo simulation that predicts the vertical
velocity dispersion of brown dwarfs  in the Milky Way, using realistic
cooling  expectations.   We  link  our  velocity  dispersions  to  the
vertical scale height of the  brown dwarf population, which we predict
is $\sim\!15$\%  thinner than  that for T0  dwarfs. We  recognize that
this decrease is  due to atmospheric cooling in the  brown dwarfs, and
the spectral type of the departure from constancy would signal the the
average  hydrogen-burning  limit or  where  cooling  is a  significant
factor.   Given the  current sample  sizes and  survey conditions,  we
speculate this may be the  purview of future wide-field missions (such
as WFIRST, LSST, or Euclid).

\appendix

\section{The Effect of the Lower Mass Limit}\label{sec:masslimits}

\begin{figure}[ht]
%  \epsscale{1}
%  \plotone{temp_m0.ps}
  \begin{center}
  \includegraphics[width=0.8\textwidth]{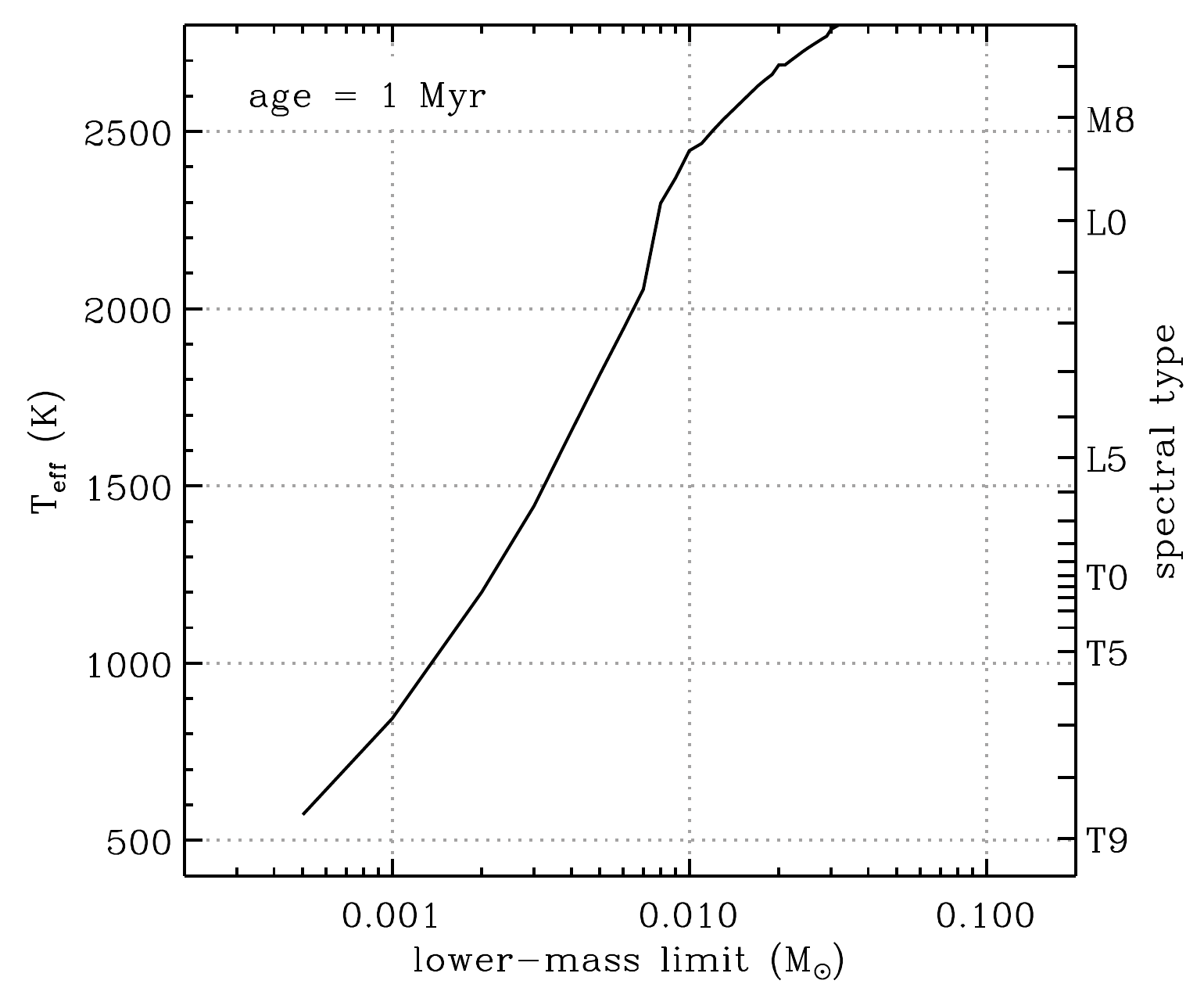}
  \end{center}
  \caption{The   effective    temperature   at    $a\!=\!1$~Myr   from
    \citet{burrows97}  with  spectral types  from  \citet{filippazzo}.
    The  velocity  dispersions  for  the non-cooling  models  will  be
    incomplete for spectral types when  the lower-mass limit is larger
    than this  relation.  For  example, the spectral  type T5  will be
    incomplete    if    the    lower-mass     limit    is    set    as
    $m_0\!\geq\!0.0015$~M$_{\odot}$.  However the  cooling models will
    only suffer this  incompleteness if there is  a significant amount
    of recent  star formation.  As discussed  in \App{sec:masslimits},
    the incomplete velocity dispersions will  be biased to older stars
    and   be  larger   than  that   of  the   complete  types.    Upon
    experimentation  with  the  lower-mass  limit, we  find  that  our
    velocity dispersions are not  substantially affected by our choice
    of $m_0\!=\!0.0005$~M$_{\odot}$.
    \label{fig:m0}}
\end{figure}

The lower-mass limit roughly sets the  the range of spectral types for
which  our  simulation  is  complete.  Therefore,  we  considered  our
nominal                Galaxy                 model                for
$(\alpha,\beta,\gamma)\!=\!(0.5,4.5,0.7~\mathrm{Gyr})$ with a range of
lower-mass                          limits                          of
$m_0\!\in\![0.0005,0.001,0.0015,0.002,0.0025]$~M$_{\odot}$.  Obviously
the non-cooling  models are profoundly affected:  the simulation never
creates  stars   below  a  certain  temperature,   since  the  initial
temperatures are a fixed function  of stellar mass (see \Fig{fig:m0}).
The cooling  models are  mostly unaffected  by varying  the lower-mass
limit  since the  majority of  the  star formation  occurred at  early
times. If  we consider SFHs with  more recent star formation  (such as
$\tau\!\sim\!t_0$, where $t_0$ is the age of the Milky Way disk), then
the velocity dispersions of T-dwarfs  become larger than those for the
L-dwarfs, and this disparity  increases with increasing the lower-mass
limit.   This arises  because  the  only valid  T-dwarfs  must be  old
(massive), and  have cooled  to their lower  temperatures ---  and the
young  (low mass)  dwarfs were  artificially omitted.   Therefore SFHs
with significant  recent star formation  may be incomplete  in T-dwarf
velocity dispersions if the lower-mass limit is too large.  As a final
note, this effect becomes even  more exacerbated for very bottom-heavy
IMFs, where  the relative  number of stars  near the  lower-mass limit
becomes large.

\section{Analytic Integration}\label{sec:integ}

In the  above, we presented  a Monte Carlo calculation  for predicting
the vertical  velocity dispersion as  a function of spectral  type (or
effective temperature).  We pursued this avenue for analysis so we may
make    direct    comparisons    to    various    published    results
\citep[e.g.][]{burg04,deacon06,dayjones13,burg15,marocco15},   however
it is worth mentioning that these Monte Carlo simulations are simply a
way of numerical  approximating a complex integral.   We have verified
that  our  simulations  reproduce  those published  results,  but  are
exactly given by:
\begin{equation} \label{eqn:sigma}
\left<\sigma^2_w\right>=\frac{\int\int\psi(t)\,\phi(m)\,\sigma^2_w(t_0-t)\,\Pi(T_{eff}(t_0-t,m);T_0,T_1)\,\dd t\,\dd m}{\int\int\psi(t)\,\phi(m)\,\Pi(T_{eff}(t_0-t,m);T_0,T_1)\,\dd t\,\dd m}
\end{equation}
where $\Pi(\cdot)$ is the boxcar-windowing function:
\begin{equation}
  \Pi(x;x_0,x_1)=\begin{cases}
    0 & x < x_0\\
    1 & x_0 \leq x \leq x_1\\
    0 & x_1 < x,
  \end{cases}
\end{equation}
and  $T_0$/$T_1$ are  the minimum/maximum  effective temperature  that
correspond  to  a given  maximum/minimum  spectral  type in  question,
respectively.  \Eqn{eqn:sigma} is similar  to the results described by
\citet{ab09}, but extended to include atmospheric cooling.

\acknowledgments We would  like to thank the  Referee, Adam Burgasser,
for  the  great comments  and  advice.   We acknowledge  many  helpful
discussions with Pat Boeshaar, Stefano  Casertano, and Tony Tyson that
were indispensable in preparing this  work.  We also grateful to Casey
Papovich for  providing the star-formation histories  of typical Milky
Way-like galaxies.  Support  for program \#13266 was  provided by NASA
through a grant  from the Space Telescope Science  Institute, which is
operated  by   the  Associations  of  Universities   for  Research  in
Astronomy,  Incorporated,  under  NASA   contract  NAS  5-26555.   RAW
acknowledges  NASA JWST  grants NAG5-12460  and NNX14AN10G  from GSFC.
JIL acknowledges support from the JWST Project: NNX12AK01G from GSFC.

\end{document}